\documentclass[aps,prb,preprint,groupedaddress,showpacs]{revtex4-1}
\usepackage[pdftex]{graphicx,graphics}

\usepackage[T1]{fontenc}
\usepackage[latin1]{inputenc}
\usepackage[american]{babel}
\usepackage[outdir=./]{epstopdf}
\usepackage{bm}
\usepackage{natbib}
\usepackage{citesort}
\usepackage{amssymb}
\usepackage{amsmath}
\usepackage{gensymb}
\usepackage{textcomp,gensymb}

\usepackage{afterpage}
\usepackage{capt-of}% or use the larger `caption` package

\usepackage{rotating}% 
\usepackage{lipsum}% dummy text
\usepackage{braket}

\usepackage{tabularx}

\usepackage{color}

\usepackage{mathptmx}

\begin{document}

\title{Antiferromagnetic exchange weakening in the TbRhIn$_5$ intermetallic system with Y-substitution.}

\author{R. P. Amaral}
\affiliation{Instituto de F\'isica, Universidade Federal de Uberl\^andia, 38408-100, Uberl\^andia, MG, Brazil}

\author{R. Lora-Serrano}
\email{rloraserrano@ufu.br}
\affiliation{Instituto de F\'isica, Universidade Federal de Uberl\^andia, 38408-100, Uberl\^andia, MG, Brazil}

\author{W. A. Iwamoto}
\affiliation{Instituto de F\'isica, Universidade Federal de Uberl\^andia, 38408-100, Uberl\^andia, MG, Brazil}

\author{D. J. Garcia}
\affiliation{Consejo Nacional de Investigaciones Cient\'ificas y T\'ecnicas (CONICET) and Centro At\'omico Bariloche, S.C. de Bariloche, R\'io Negro, Argentina}

\author{D. Betancourth}
\affiliation{Centro At\'omico Bariloche and Instituto Balseiro (U. N. Cuyo), 8400 Bariloche, R\'io Negro, Argentina}

\author{J. M. Cadogan}
\affiliation{School of Physical, Environmental and Mathematical Sciences,
UNSW Canberra at the Australian Defence Force Academy, Canberra, ACT, BC 2610, Australia}

\author{S. Mu\~{n}oz-P\'{e}rez}
\affiliation{School of Physical, Environmental and Mathematical Sciences,
UNSW Canberra at the Australian Defence Force Academy,
Canberra, ACT, BC 2610, Australia}

\author{M. Avdeev}
\affiliation{Bragg Institute, ANSTO, PMB 1, Menai, NSW 2234, Australia}

\author{R. Cobas-Acosta}
\affiliation{School of Physical, Environmental and Mathematical Sciences,
UNSW Canberra at the Australian Defence Force Academy,
Canberra, ACT, BC 2610, Australia}

\author{E. M. Bittar}
\affiliation{Centro Brasileiro de Pesquisas F\'isicas, Rua Dr. Xavier Sigaud 150, 22290-180 Rio de Janeiro, RJ, Brazil}

\author{J. G. S. Duque}
\affiliation{Programa de P\'os-Gradua\c{c}\~ao em F\'isica, Campus Prof. Jos\'e Alu\'isio de Campos, UFS, 49100-000 S\~ao Crist\'ov\~ao, SE, Brazil}

\author{P. G. Pagliuso}
\affiliation{Instituto de F\'isica ``Gleb Wataghin'', UNICAMP, 13083-970, Campinas-S\~ao Paulo, Brazil}

\date{\today}

%\maketitle

%\keywords{Mean-field model, crystalline electric field, specific-heat, magnetic susceptibility}

\begin{abstract}

%\noindent 

We report measurements of the temperature dependence specific heat, magnetic susceptibility in single crystals of the series of intermetallic compounds Tb$_{1-x}$Y$_x$RhIn$_5$ (nominal concentrations $x= 0.0, 0.15, 0.3, 0.4, 0.5, 0.7$).  A mean field approximation to simulate the macroscopic properties along the series has been used. Neutron diffraction data in powdered samples of Tb$_{0.6}$Y$_{0.4}$RhIn$_5$ and Tb$_{0.6}$Y$_{0.4}$RhIn$_5$ reveal antiferromagnetic (AFM) propagation vector $k=[\frac{1}{2}~0~\frac{1}{2}]$ with the magnetic moments oriented along the tetragonal \textit{c} axis or canted from the \textit{c}-axis, respectively for Y and La-substitutions. Considering both the simulations of the magnetic exchange and neutron diffraction data, we discuss the role of combined effects of crystalline electric field (CEF) perturbations and dilution in the evolution of magnetic properties with Y and La contents. In particular, we found negligible variations of the $B^{m}_{n}$ parameters along the Y series. The decrease of $T_N$ with \textit{x} is fully dominated by magnetic dilution effects.

\end{abstract}


\maketitle

\section{Introduction}

Interesting ground states (GS) can be observed in condensed matter by tuning physical properties with chemical substitutions. This is particularly true in strongly correlated 4\textit{f}-electrons systems where many of the observed phenomena include unconventional superconductivity, complex magnetic order, quantum criticality, heavy-fermion behavior, magnetic transitions, among others\cite{Lohneysen2007,SteglichScience2010,review}. The occurrence of each of these GS depends on the hybridization between 4\textit{f} electrons with the conduction electrons \cite{Doniach}. In this context, intermetallic compounds from the Ce$_mM_n$In$_{3m+2n}$ family ($M=$Co, Rh, Ir) have became one of the important attractions to understand the effects of doping in tuning low energy states such as antiferromagnetism (AFM), unconventional superconductivity (USC), Non-Fermi-liquid behavior (NFL) and Kondo effect\cite{pagliuso1,pagliuso2,Fisk5,Hering,bauer2,Zapf,Moreno1,pagliuso7,VictorCorrea,Light,Alver,christianson2}. 
In the last fifteen years, many dilution studies in the above series were conducted for both ambient pressure AFM (CeRhIn$_5$) and USC (CeCoIn$_5$) heavy-fermion compounds\cite{pagliuso7,VictorCorrea,Light,Alver,christianson2,wei3,Tanatar,Petrovic3,Satoro}.

In general, the magnetic properties of non-Ce isostructural related compounds from the above family depend on the localized character of \textit{f} electrons. This has proved useful in the systematic study of the  dimensionality and/or anisotropy effects influence on the GS of their members.

Searching a complete microscopic understanding along the $R_mM_n$In$_{3m+2n}$ family (\textit{R}: rare earth), here we studied the evolution of $4f$-electrons magnetism along structurally-related compounds with $R=$ Tb. Recently, we demonstrated that diluting with non-magnetic Lanthanum in the antiferromagnetic TbRhIn$_5$ decreases Néel temperature with a non-linear behaviour as a function of La concentration and extrapolates to zero at roughly 70\%
of La content (dilution limit)\cite{raimundo5}, differently from the observed $\sim$40\% for Ce$_{1-x}$La$_{x}$RhIn$_5$, 
(Ce$_{1-x}$La$_{x})_2$RhIn$_8$ or Nd$_{1-x}$La$_{x}$RhIn$_5$ families\cite{raimundo7}. This has been related to the competing CEF effects, Tb-Tb exchange and disorder\cite{raimundo5}. Furthermore, La-dilution in the \textit{S} system Gd$_{1-x}$La$_{x}$RhIn$_5$ (negligible CEF effects) proved to induces substitutional disorder with a distribution of critical temperatures as a function of \textit{x}\cite{raimundo8} and confirmed the relevance of CEF effects on the magnetic properties of Tb and Nd-based \textit{R}RhIn$_5$ compounds studied in refs. \onlinecite{raimundo2,raimundo5,raimundo7}. Gd$_{1-x}$La$_{x}$RhIn$_5$ represents then a simple 4\textit{f} ($L=$0) AFM system for the study of substitutional disorder effects and short range order in antiferromagnets.

In this manuscript we conducted a systematic study on the non-magnetic Y-substitution in the series Tb$_{1-x}$Y$_{x}$RhIn$_5$  for nominal concentrations $x = 0.15, 0.3, 0.4, 0.5, 0.7$ and $1.0$. 
The non substituted TbRhIn$_5$ ($x=$ 0) orders antiferromagnetically below $T_N$ $\sim$ 46 K with a commensurate propagation vetor\cite{raimundo2}. Y$^{3+}$ has approximately the same ionic size as Tb$^{3+}$ at the 1\textit{a} site (twelve In atoms as first neighbours), thus it is expected the effect of dilution be roughly the same as in the case of La$^{3+}$,\cite{raimundo5} but the chemical pressure is practically non existent for Tb$^{3+}$ with Y-substitution. Therefore, comparative studies between doping with Y$^{3+}$ and La$^{3+}$ may isolate the effects of chemical pressure from those of CEF variation when using chemical dilution as a control variable. In this work, the N\'eel temperature is suppressed with increasing Yttrium less drastically than when Tb$^{3+}$ is substituted by La$^{3+}$. Dilution by weakening the magnetic exchange interactions between rare earth ions should be far more important than perturbations from the crystalline potential (i.e. CEF effects) when Tb$^{3+}$ is substituted by Y$^{3+}$. Neutron diffraction data in the Tb$_{0.60}$Y$_{0.40}$RhIn$_5$ sample reveal the magnetic propagation vector [$\frac{1}{2}$,0,$\frac{1}{2}$] with magnetic moments oriented parallel to the tetragonal \textit{c}-axis. It is the same propagation vector as obtained for TbRhIn$_5$ and, together with the results from a model to simulate CEF and exchange interactions in these systems, it suggests negligible differences of the CEF scheme along the series and a mean field behaviour of the main $J_{RKKY}$'s exchange parameters.

\section{Experimental}

The single-crystalline samples of  Tb$_{1-x}$Y$_{x}$RhIn$_{5}$ were grown by the metallic Indium (In) excess flux method.\cite{Fisk2,Fisk3} High purity Terbium (4N), Yttrium (4N), Rh(3N) and In (4N) in the proportion (1-x):x:1:20 were put in an Alumina crucible and sealed with vacuum of $10^{-2}$  Torr in quartz tube. Six compositions with nominal concentrations x = 0.15, 0.3, 0.4, 0.5, 0.7 and 1.0 were prepared and studied. The non magnetic x = 1.0 sample was synthesized in order to extract the phonon contribution to the specific heat data of doped samples, as well as used for comparing crystallographic data along the series. Crystals grow with a platelet-like morphology, and the tetragonal $[001]$ direction perpendicular to the macroscopically observed \textit{ab}-plane. This is usually confirmed by Laue diffraction data. 

Room temperature (RT) X-ray powder diffraction (XRD), in the Bragg-Brentano geometry, graphite monochromator and Cu K$_{\alpha}$ radiation, allows for checking the formation of the tetragonal HoCoGa$_5$-type structure (space group $P4/mmm$). The measurements were done over a scattering angle 2$\theta$ from 10 to 110$^{\text{o}}$, with a step of 0.02$^{\text{o}}$. In order to determine structural parameters along the series, the XRD data were least-squares Rietveld refined using the FullProf/WinPlotr software package.\cite{fullprof,winplotr}  
The actual Yttrium concentration was measured with Energy-dispersive X-ray spectroscopy (EDS) data taken in a Thermo Scientific Noran System 7, attached to a JEOL JSM-6490LV scanning electron microscope, accelerating voltage of 30 keV and NanoTrace detector. Temperature dependent magnetic susceptibility, after zero-field cooling, and specific heat data were collected on a commercial Quantum Design PPMS. 
The electrical resistance was measured using the PPMS low-frequency \textit{ac} resistance bridge and four-contact configuration. The single crystal samples used in the electrical resistance measurements were screened to be free of surface contamination by residual Indium flux. 
Powder neutron diffraction (PND) experiments were carried out on the Echidna high-resolution powder diffractometer at the OPAL reactor in Sydney, Australia. The neutron wavelength used was $2.4395$\AA. Because the Rh and In are both fairly strong absorbers, each pattern was counted for 12 hours.

\section{Results and Analysis}

\subsection{X-ray powder diffraction}

\begin{figure}[h!]
	\begin{center}
		\includegraphics[%
		width=0.9\linewidth,
		keepaspectratio]{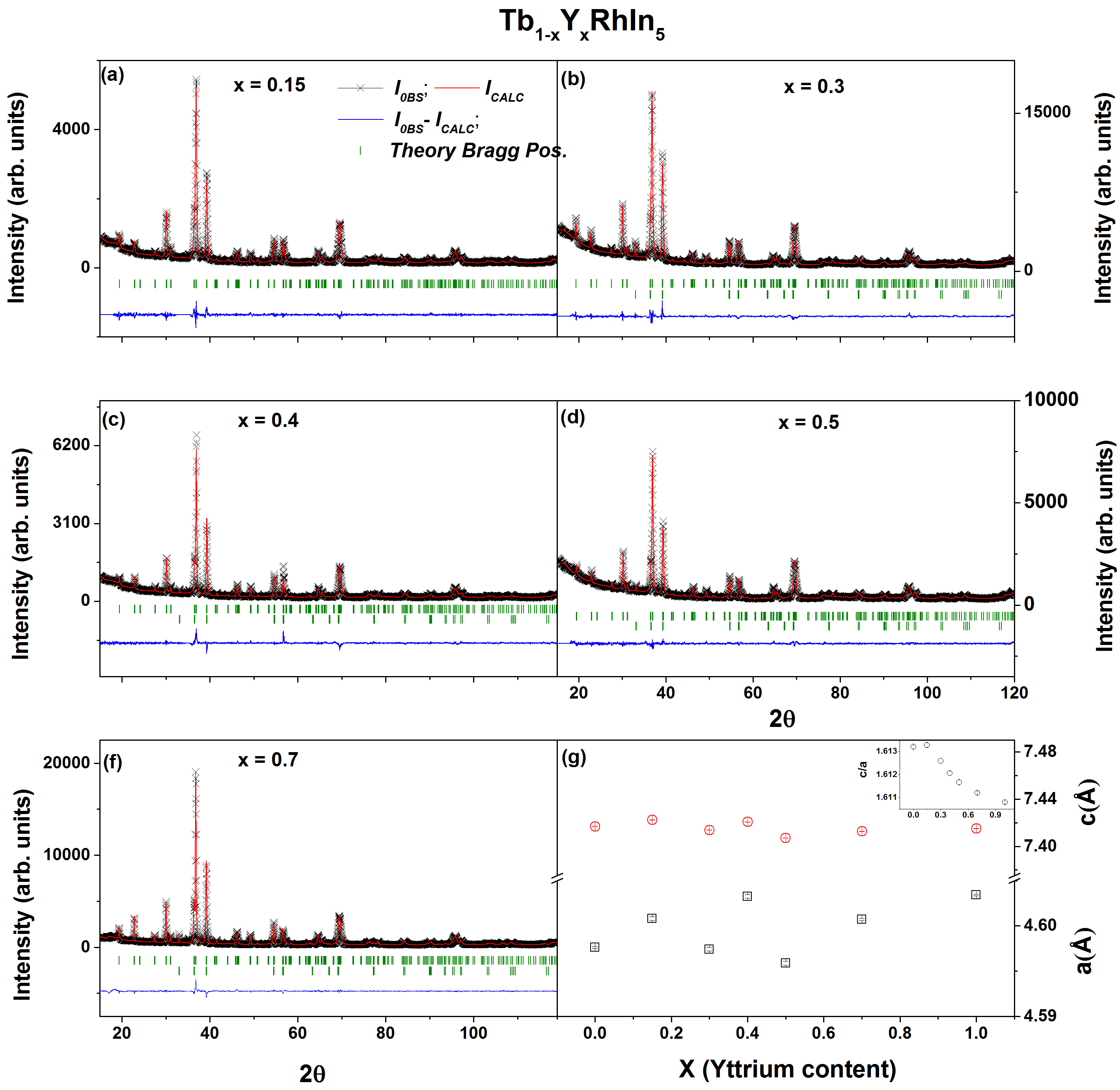}
	\end{center}
	\caption{Rietveld refinement of the X-ray powder diffraction data of Tb substituted compositions, from (a) \textit{x}=0.15 to (f) \textit{x}=0.7; observed ($I_{OBS}$), calculated ($I_{CALC}$) and difference ($I_{OBS}-I_{CALC}$) data are shown, vertical bars represent the theoretical Bragg positions for main 1-1-5 phase and the secondary tetragonal Indium $I4/mmm$ phase. Panel (g) is the tetragonal lattice cell parameters $a$ and $c$ vs.~Yttrium concentration in Tb$_{1-x}$Y$_x$RhIn$_5$; inset in is the $c/a$ ratio vs. $x$.}
	\label{FigCellParam}
\end{figure}

For the Rietveld refinement, the starting model used was the structure of HoCoGa$_5$\cite{YGrin} with cell parameters from [\onlinecite{raimundo2}]. As a result of the flux method, Indium excess remains in the crystal surfaces and its Bragg reflections can be observed in the XRD data. Therefore, its contribution was initially excluded from the refinement, but in those XRD data where the reliability factors improved remarkably with its inclusion, it was considered as a second phase at the final stages. Tb and Y ions were allowed to share the 1\textit{a} position adding to a full site occupancy and then refined. In2 \textit{z} coordinate and isotropic thermal (displacement) parameters were also refined. The refinement of In2 \textit{z} improved the residual factors for all \textit{x}; thermal \textit{B}'s did not improve the quality of the calculations and were kept constants. Figs. \ref{FigCellParam}(a)-(f) show the Rietveld refinement results from the substituted compounds (\textit{x}=0.15 to \textit{x}=0.7). The legend in (a) details the symbols used for each observed ($I_{OBS}$), calculated ($I_{CALC}$) and difference ($I_{OBS}-I_{CALC}$) data in all panels, as well as the theoretical Bragg positions (vertical bars) for both phases. Panel (g) is the evolution of the tetragonal cell parameters $a$ and $c$ along the series for nominal $x=0.0-1.0$. Error bars are mostly smaller than the symbols used and cannot be observed. Inset is the \textit{c/a} ratio showing a slight decrease up to $x=$1.0. From Fig. \ref{FigCellParam}(g) the unit cell size does not change significantly. This is relevant in the context of evolution of the crystalline electric potential with \textit{x} and will be discussed below. On the other hand, the decrease in the $c/a$ ratio could be an indication of local changes without altering the unit cell volume (not shown).

\begin{table}[h!]
	\centering
	%\resizebox{\columnwidth}{!}{%
	\scalebox{0.8}{	
		\begin{tabular}{|c|c|c|c|c|c|c|c|} \hline \hline
			$x$ & $x=$0.0 & $x=$0.15 & $x=$0.3 & $x=$0.4 & $x=$0.5 & $x=$0.7 & $x=$1.0  \\ \hline \hline
			$R_p$    & 5.63    & 5.58 & 3.69 & 4.04 & 5.85 & 4.81 & 5.39  \\ \hline
			$R_wp$   & 7.51    & 7.15 & 4.71 & 5.13 & 7.55 & 6.14 & 6.73  \\ \hline
			$\chi^2$ & 1.53    & 1.44 & 2.68 & 1.55 & 2.09 & 2.40 & 2.12  \\ \hline
			\multicolumn{8}{|c|}{Unit cell parameters (\AA)}  \\ \hline
			\textit{a} (\AA)       & 4.59761(13) & 4.60082(17) & 4.59742(16) & 4.60325(16) & 4.60071(9) & 4.60069(11) & 4.60340(8)  \\ 
			\textit{c}(\AA)        & 7.4169(3) & 7.4224(3) & 7.4138(4) & 7.4207(3) & 7.4157(3) & 7.41270(20) & 7.41510(20)  \\ \hline
			\multicolumn{8}{|c|}{(Tb/Y)In$_3$ cuboctahedra interatomic distances (\AA)} \\ \hline
			(Tb/Y)-In1$\times4$         & 3.25100(7) & 3.25327(9) & 3.25087(8) & 3.25499(8) &3.25319(5) & 3.25318(6) & 3.25510(4)  \\
			(Tb/Y)-In2$\times8$          & 3.2058(6)  & 3.20650(9) & 3.2077(8) & 3.2117(9) & 3.2057(10) & 3.2029(6) & 3.2082(8)  \\ \hline
			\multicolumn{8}{|c|}{Angles ($^{\text{o}}$)}   \\ \hline
			In1-(Tb/Y)-In1 & 90.000(4) & 90.000(5) & 90.000(4)   & 90.000(4) & 90.000(3)  & 90.000(3) & 90.000(4) \\ \hline
			In1-(Tb/Y)-In2 & 59.532(16) & 59.516(3) & 59.55(2) & 59.55(2) & 59.51(3) &	59.480(16) & 59.515(20)  \\ \hline
			In1-(Tb/Y)-In2 & 120.468(16) & 120.484(5) & 120.45(2) & 120.45(2) & 120.49(3) & 120.520(16) & 120.485(2)  \\ \hline
			In2-(Tb/Y)-In2 & 91.628(16) & 91.684(5) & 91.55(2) & 91.55(2) & 91.71(3) & 91.813(16) & 91.689(2) \\ \hline
			In2-(Tb/Y)-In2 & 60.935(16) & 60.967(3) & 60.89(2) & 60.89(2) & 60.98(3) & 61.041(16) & 60.97(2) \\ \hline
			In2-(Tb/Y)-In2 & 88.37(4) & 88.316(5) & 88.45(5) & 88.45(5) & 88.29(6) & 88.19(4) & 88.31(4) \\ \hline
			\multicolumn{8}{|c|}{RhIn$_2$ parallelepipeds interatomic distances (\AA)} \\ \hline
			(Rh)-In2$\times8$     & 2.7308(5)  & 2.73399(8) & 2.7284(6) & 2.7312(7) & 2.7327(8) & 2.7341(5) & 2.7325(6) \\ \hline
			\multicolumn{8}{|c|}{Angles ($^{\text{o}}$)}   \\ \hline
			In2-Rh-In2 & 73.060(14) & 73.021(3)  & 73.132(19) & 73.15(2) & 73.06(2) & 73.016(14)  & 73.114(18) \\ \hline
			In2-Rh-In2  & 65.34(3)  & 65.421(4) & 65.19(4) & 65.14(5) & 65.34(5) & 65.43(3) & 65.23(4) \\ \hline \hline
		\end{tabular}%
%	}
}
	\caption{Rietveld refinement reliability parameters, unit cell dimensions and main interatomic distances (in \AA) and angles for the series Tb$_{1-x}$Y$_{x}$RhIn$_5$.}
	\label{interatomicdistances}
\end{table}

Selected interatomic distances and angles extracted from the refinements of XRD data are given in Table \ref{interatomicdistances}.

\subsection{Chemical analysis}

\begin{figure}[h!]
	\begin{center}
		\includegraphics[%
		width=0.85\linewidth,
		keepaspectratio]{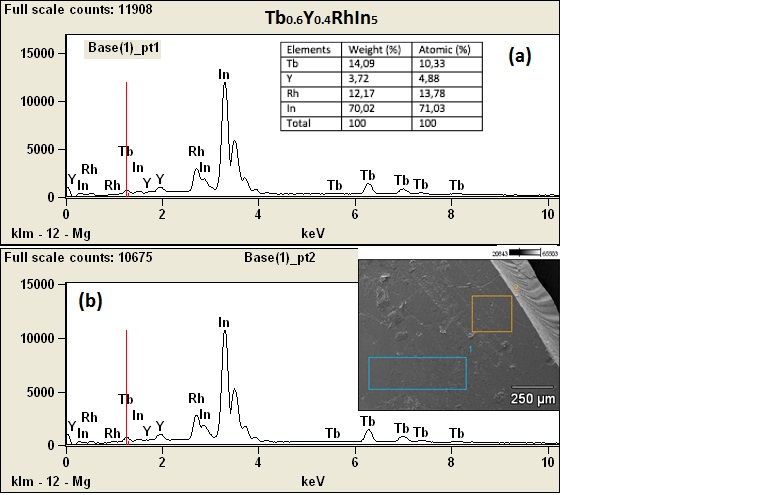}
	\end{center}
	\caption{EDS spectrum of the x = 0.4 (nominal) sample for two different regions of the single crystal [(a) and (b)]. The operated areas are shown in the inset of (b).}
	\label{FigEDS}
\end{figure}

\begin{table}
	\centering
%	\resizebox{\columnwidth}{!}{%
	\scalebox{0.9}{	
		\begin{tabular}{cc} \hline \hline
			 Nominal composition & Chemical composition \\ \hline
			 Tb$_{0.85}$Y$_{0.15}$RhIn$_5$ &  Tb$_{0.95}$Y$_{0.05}$Rh$_{0.96}$In$_{4.98}$  \\
			  Tb$_{0.7}$Y$_{0.3}$RhIn$_5$ &  Tb$_{0.88}$Y$_{0.12}$Rh$_{0.96}$In$_{5.02}$  \\
			 Tb$_{0.6}$Y$_{0.4}$RhIn$_5$ &  Tb$_{0.7}$Y$_{0.3}$Rh$_{0.96}$In$_{4.97}$ (Fig. 2)  \\
			 Tb$_{0.5}$Y$_{0.5}$RhIn$_5$ &  Tb$_{0.65}$Y$_{0.35}$Rh$_{0.96}$In$_{4.95}$  \\
			 Tb$_{0.3}$Y$_{0.7}$RhIn$_5$ &  Tb$_{0.38}$Y$_{0.62}$Rh$_{0.97}$In$_{4.90}$  \\  \hline \hline
		\end{tabular}%
	}
	\caption{Nominal and chemical compositions along the Tb$_{1-x}$Y$_{x}$RhIn$_5$ series. Data were obtained by averaging over two available regions.}
	\label{EDS}
\end{table}

EDS data were collected to confirm the elements content of the substituted samples. Fig.~\ref{FigEDS} shows representative patterns of the Tb$_{0.6}$Y$_{0.4}$RhIn$_5$ sample. The average quantitative atomic and weight percentage of the compositional elements are indicated in the inset of Fig.~\ref{FigEDS}(a). Results in Figs.~\ref{FigEDS}(a) and (b) were obtained from two investigated regions [inset of (b)]. The same procedure were conducted for all the studied compositions. From these analysis, the crystals contain Tb, Y, Rh and In elements in the average chemical compositions of Table \ref{EDS}. In particular, from the data of Fig.~\ref{FigEDS} it is obtained a Tb content of around 0.7, which agrees with the neutron diffraction data below. All the actual/chemical compositions were confirmed by the magnitude of the effective paramagnetic Tb$^{3+}$ moment extracted from the linear fit to the high temperature region of the inverse susceptibility data. Despite this shift from the nominal content we did not observe the presence of intrinsic secondary phases different from the 1-1-5-type structure. Indeed, the occupancy parameters in the Rietveld refinement above were fixed to the chemical compositions observed. 
Therefore, all the data presented and discussed below are in terms of chemical/actual Y concentrations.

\subsection{Magnetic characterisation}

\begin{figure}[h!]
	\begin{center}
		\includegraphics[%
		width=0.55\linewidth,
		keepaspectratio]{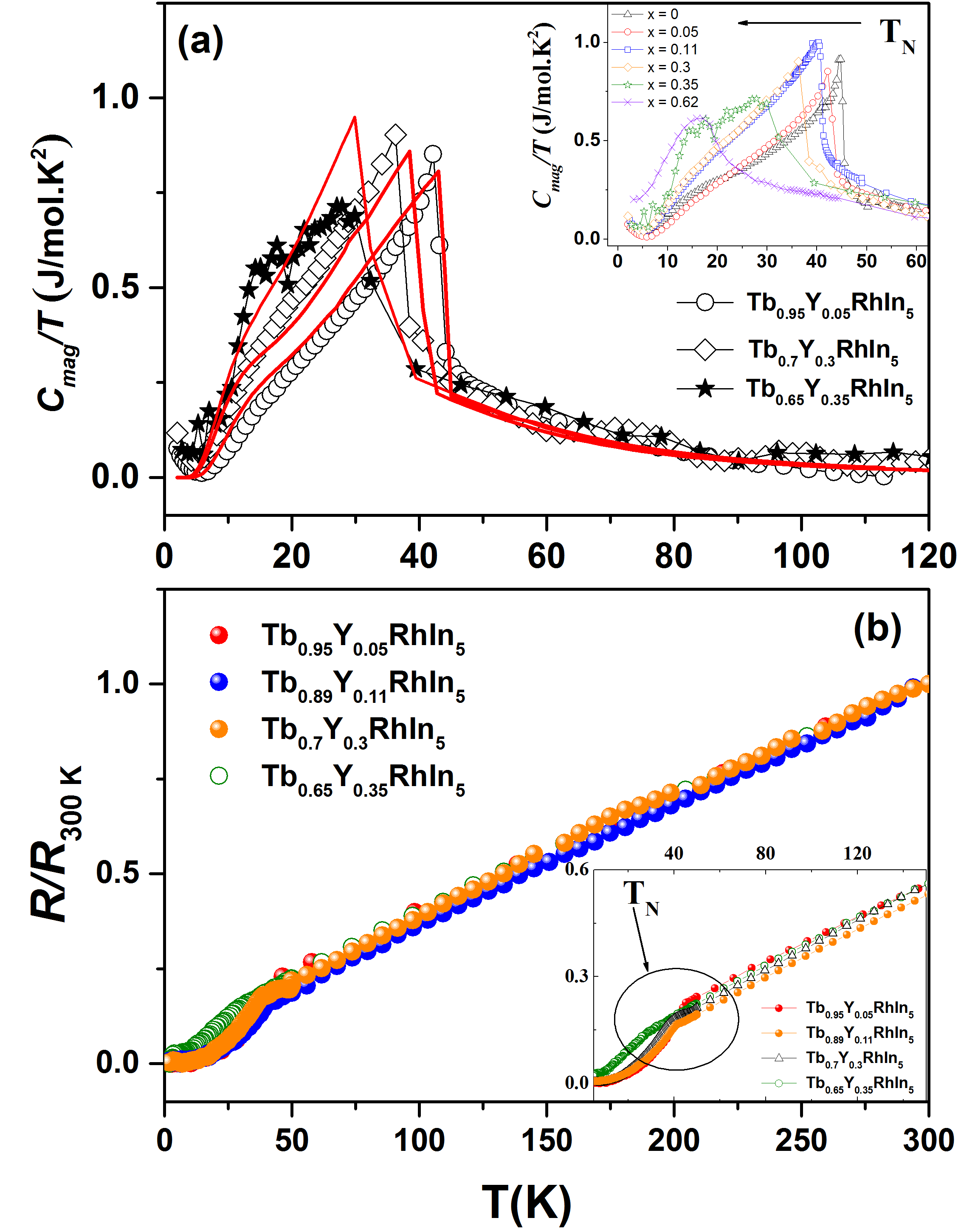}
	\end{center}
	\caption{(Color online) (a) Magnetic contribution to the specific heat divided by temperature for representative concentrations in  Tb$_{1-x}$Y$_x$RhIn$_5$ (\textit{x} = 0.05, 0.28, 0.35). Continuous curves are fits to the data by using the model described below. The inset shows a closer view of the evolution of the Lambda anomaly with \textit{x}. (b) Temperature dependence of the electrical resistance for $x=0.05-0.35$, normalized by the resistance at $T=300 K$, $R/R_{300K}$. The inset highlights the anomaly around $T_N$, which coincides with the $T_N$ as extracted from the inflection point of heat capacity data.}
	\label{FigCpRho}
\end{figure}

Fig.~\ref{FigCpRho}(a) shows the temperature evolution (2$<T<$120 K) of magnetic heat-capacity $C_{mag}(T)/T$ for the Tb$_{1-x}$Y$_x$RhIn$_5$ series. For extracting $C_{mag}/T$, the lattice specific heat was estimated from the non-magnetic data of YRhIn$_{5}$ (\textit{x} = 1) and subtracted from the total specific heat. The inflection points of the $C_{mag}(T)/T$ vs. \textit{T} curves have been defined as $T_N$. The metallic character of the samples were tested by measuring the temperature dependence of the normalized electrical resistance, $R/R_{300K}$, for $x=0.05-0.35$ (Fig.~\ref{FigCpRho}(a)). $R_{300K}$ stands for the resistance taken at RT and all data were collected at \textit{H} = 0 and applied \textit{dc} current along the \textit{ab}-plane. As for TbRhIn$_5$,\cite{raimundo2} substituted compounds exhibit a typical metallic (linear) behavior above 50 K, while a clear kink can be seen at $T_N$.

\subsection{Crystalline field excitations with Y-content}
\label{CEF}

The evolution of the crystal-field ground-state configurations have been accompanied by using a mean field model
including anisotropic first-neighbours RKKY interaction and the tetragonal CEF Hamiltonian. The \textit{f}-electron magnetism in these series can be studied with the Hamiltonian:\cite{pagliuso5,DuqueRNi3Ga9}

\begin{equation}
H = H_{CEF} - \sum_{i, k}{j_{ik} \textbf{J}_{i} \cdot \textbf{J}_{k}} - g \mu_B \textbf{H}_{0} \cdot \sum_{i}{\textbf{J}_i};
\label{hamiltonian}
\end{equation}

the second term to the right is the magnetic interaction between the $J_i$ and $J_k$ moments. $j_{ik} = j_0, j_1, j_2, j_3$ and $j_4$ for first and second nearest rare earth neighbors along the tetragonal [100], [110], [001], [101] and [111] directions, respectively. It is worth noticed that different from the effective isotropic exchange interaction used in refs.~\onlinecite{pagliuso5,raimundo2,raimundo5,raimundo7}, here we consider an anisotropic exchange between \textit{R} ions along those three directions. The third term represents the Zeemann effect with an applied field $\textbf{H}_0$. The first term is the CEF Hamiltonian and it is defined as:\cite{hutchings}

\begin{equation} 
H_{CEF} = \sum_{i,n,m}{ B_{n}^{m}(i) O_{n}^{m}(i)} = {B_{2}^{0}O_{2}^{0}(i) + B_{4}^{0}O_{4}^{0}(i) + B_{4}^{4}O_{4}^{4}(i) + B_{6}^{0}O_{6}^{0}(i) + B_{6}^{4}O_{6}^{4}(i)};
\label{eqhcef}
\end{equation} 

where $O_{n}^{m}$ are the Stevens equivalent operators (they describe the CEF in terms of powers of the local total angular momentum \textbf{\textit{J}}). $B_{n}^{m}$ characterize the crystal field and can be obtained by fitting experimental data of magnetic susceptibility and specific heat (below). The $j_{ik} = j_0, j_1$ and $j_2$ anisotropic exchanges follow the notation previously used in refs.~\onlinecite{Takeuchi2006,Takeuchi2007,raimundo8} for this magnetic unit cell symmetry. $j_3$ and $j_4$ are introduced in this work as the Tb-Tb exchanges along the face and body diagonals, respectively.

\begin{figure}[h!]
	\begin{center}
		\includegraphics[%
		width=0.7\linewidth,
		keepaspectratio]{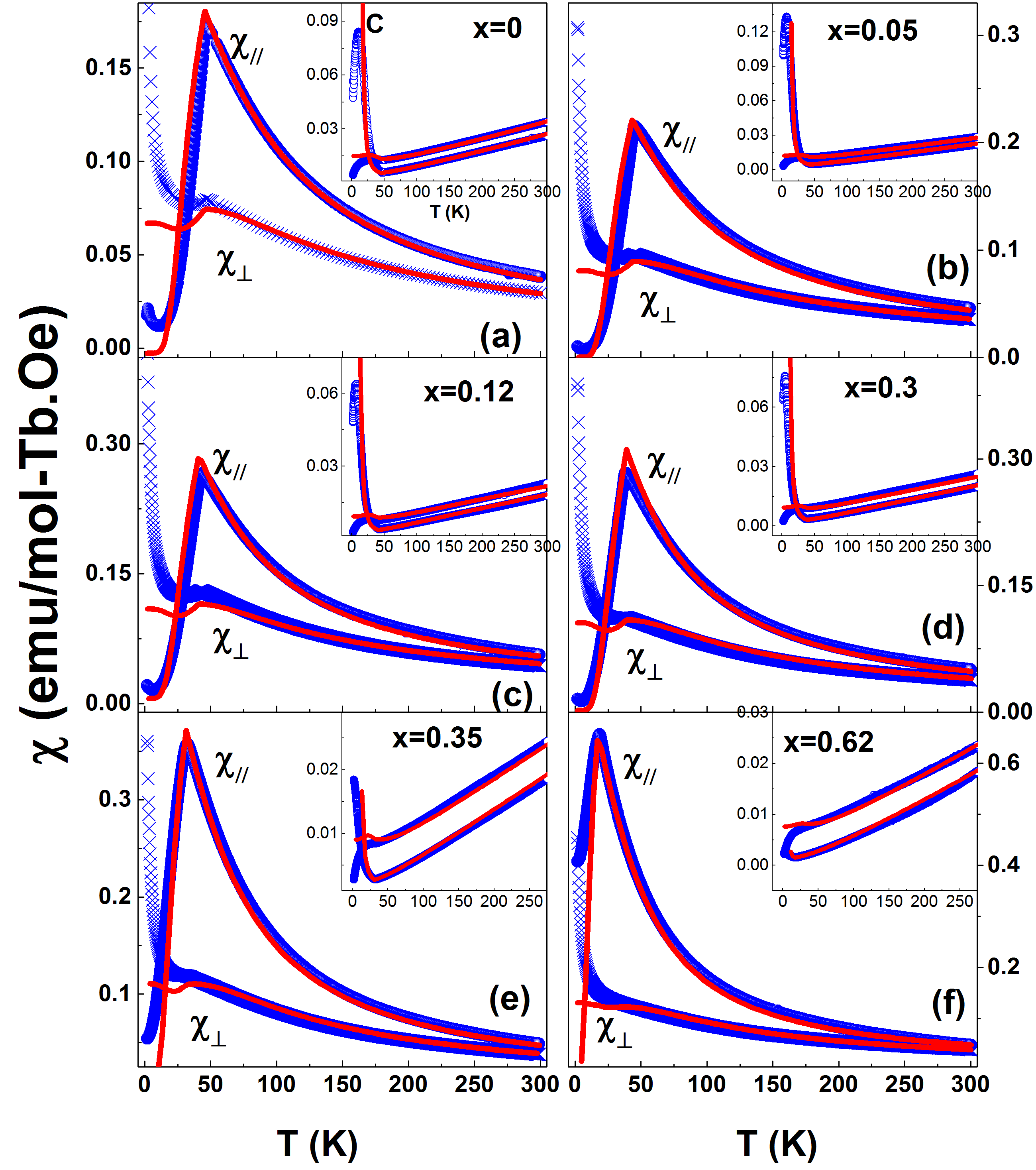}
	\end{center}
	\caption{(a)-(f) Main panels: observed temperature dependences of the magnetic susceptibility, $\chi(T)$ (scatter data), for applied field of 1 kOe parallel ($\chi_{\parallel}$) and perpendicular ($\chi_{\perp}$) to the tetragonal \textit{c}-axis. Continuous curves are the best fits to the data using the mean-field model described in the text. The insets are the experimental $\chi_{\parallel}^{-1}$ vs. \textit{T} and $\chi_{\perp}^{-1}$ vs. \textit{T} together with the best results from the model. Units in the insets are (emu/mol-Tb.Oe$^{-1}$).} 
	\label{FigChiModel}
\end{figure}

Figs. \ref{FigChiModel} show the best fits, using the above model, to the temperature dependence of magnetic susceptibility (main panels) for applied field of 1 kOe parallel to the $[100]$ crystallographic direction ($\chi_{\perp}$) and along the $[001]$ direction ($\chi_{\parallel}$). Morphologically well defined single crystal unit cell directions allowed for the anisotropic $\chi(T)$ measurements to be collected. The onset of the AFM order from the high-T paramagnetic region is signaled by maxima below 46 K, which shift to lower temperatures as Yttrium content is increased. This is expected from the weakening of Tb-Tb exchange. At lower-T an anisotropic Curie-like tail was observed in the magnetic susceptibility data for all measured crystals. It can be related to the proximity of an additional magnetic phase transitions below 2 K. The same upturn is also slightly defined in the low-T $C/T(T)$ vs. \textit{T} data, however it cannot be followed by the model. Insets depict the inverses $\chi_{\perp}^{-1}$ and $\chi_{\parallel}^{-1}$ vs. T for each \textit{x} and the fittings (continuous curves). Considering these results, together with the simultaneous fits to $C_{mag}(T)/T$ data (Fig.~\ref{FigCpRho}(a)), and knowing that mean field approximations do not account for critical fluctuations near phase transitions, our results describe very well the data behavior for the studied T interval and Y concentration. The sets of parameters extracted from the fits are presented below.

Tables \ref{resultados} presents experimental $T_N$ values and the fitting parameters $j_{RKKY}^{(ik)}$, $\mathrm{B}_n^m$  for each concentration. They were obtained to reproduce the experimental curves of Figs.~\ref{FigCpRho}(a) and~\ref{FigChiModel}. In what follows, we call the $j_{ik}$ parameters as $j_{RKKY}^{(ik)}$ to clearly differentiate from the angular momentum \textit{\textbf{J}}. The energy values from the splitting due to the crystal field, $E_i$, together with the eigenfunctions $\Psi_{i}$, are shown in Table \ref{eigenvectors}.

\begin{table}[h!]
\centering
\resizebox{\columnwidth}{!}{%
\begin{tabular}{ccccccccccccc} \hline \hline
$x$ &  $T_N$ (K) & $j_{RKKY}^{(0)}$ (meV) & $j_{RKKY}^{(1)}$ (meV) & $j_{RKKY}^{(2)}$ (meV) & $j_{RKKY}^{(3)}$ (meV) & $j_{RKKY}^{(4)}$ (meV) & $\mathrm{B}_2^0$ (meV) & $\mathrm{B}_4^0$ (meV) & $\mathrm{B}_4^4$ (meV) & $\mathrm{B}_6^0$ (meV) & $\mathrm{B}_6^4$ (meV) \\ \hline
0.0 & 45.55(5) & 0.0495 & 0.0190 & 0.01212 & -0.00143 & -0.003 & -0.11937 & -0.00036 & $1.3\times 10^{-5}$ & $0.63\times 10^{-5}$ & $0.02\times 10^{-5}$ \\
0.05 & 43.42(5) & 0.0443 & 0.0181 & 0.01384 & -0.0023 & -0.003 & -0.11679 & -0.00036 & $1.3\times 10^{-2}$ & $0.63\times 10^{-5}$ & $0.02\times 10^{-5}$ \\
0.12 & 40.96(5) & 0.0417 & 0.0156 & 0.00690 & -0.00333 & -0.003 & -0.10817 & -0.00036 & $1.3\times 10^{-2}$ & $0.68\times 10^{-5}$ & $0.02\times 10^{-5}$ \\
0.28 & 37.64(5) & 0.0310 & 0.0091 & 0.00427 & -0.00222 & $-9.3\times 10^{-5}$ & -0.10817 & -0.00036 & $1.3\times 10^{-2}$ & $0.7\times 10^{-5}$ & $0.02\times 10^{-5}$ \\
0.35 & 32.33(5) & 0.0267 & 0.0091 & 0.00168 & -0.00136 & $-9.3\times 10^{-5}$ & -0.10817 & -0.00036 & $1.3\times 10^{-2}$  & $0.7\times 10^{-5}$ & $0.02\times 10^{-5}$ \\ 
0.62 & 19.74(5) & 0.0164 & 0.0056 & 0.00051 & $-9.3\times 10^{-5}$ & $-9.3\times 10^{-5}$ & -0.10817 & -0.00036  &  $1.3\times 10^{-2}$ & $0.7\times 10^{-5}$ & $0.02\times 10^{-5}$ \\ 
\hline \hline
\end{tabular}%
}
\caption{Experimental $T_N$, as obtained from specific heat measurements, $j_{RKKY}^{(ik)}$ exchange and $\mathrm{B}_n^m$ CEF parameters used to reproduce $\mathrm{Tb}_{1-x}\mathrm{Y}_{x}\mathrm{RhIn}_{5}$ experimental curves (see text).}
\label{resultados}
\end{table}

\newpage
\begin{table}[h!]
	\centering
		\rotatebox{270}{
		\resizebox{\columnwidth}{!}{%
		%\scalebox{0.9}{	
		\begin{minipage}{\textheight}
			\begin{tabular}{ccccccccccccc} \hline \hline
					&\multicolumn{2}{c}{\hspace{1.5cm} x = 0}    &  \multicolumn{2}{c}{\hspace{1.5cm}x = 0.05}    &\multicolumn{2}{c}{\hspace{1.5cm} x = 0.12}    &\multicolumn{2}{c}{\hspace{1.5cm} x = 0.3}    & \multicolumn{2}{c}{\hspace{1.5cm} x = 0.35}   &\multicolumn{2}{c}{\hspace{1.5cm} x = 0.62}    \\ \hline
					Level &\hspace{1.5cm} $\Psi_{i}$ \hspace{1.5cm}             &$E_i$(K)&\hspace{1.5cm} $\Psi_{i}$ \hspace{1.5cm}             &$E_i$(K)&\hspace{1.5cm} $\Psi_{i}$ \hspace{1.5cm}              &$E_i$(K)&\hspace{1.5cm} $\Psi_{i}$ \hspace{1.5cm}              &$E_i$(K)&\hspace{1.5cm} $\Psi_{i}$ \hspace{1.5cm}              &$E_i$(K)&\hspace{1.5cm} $\Psi_{i}$ \hspace{1.5cm}             &$E_i$(K)\\ \hline
					1     &$\Ket{6}$              &$0$     &$\Ket{6}$              &$0$     &$\Ket{6}$              &$0$     &$\Ket{6}$              &$0$     &$\Ket{6}$              &$0$     &$\Ket{6}$              &$0$     \\   
					2     &$\Ket{-6}$             &$0$     &$\Ket{-6}$             &$0$     &$\Ket{-6}$             &$0$     &$\Ket{-6}$             &$0$     &$\Ket{-6}$             &$0$     &$\Ket{-6}$             &$0$     \\
					3     &$\Ket{5}$              &$45.2$  &$ \Ket{5}$             &$43.9$  &$\Ket{5}$              &$37.4$  &$\Ket{5}$              &$36.3$  &$\Ket{5}$              &$36.3$  &$\Ket{5}$              &$36.3$  \\
					4     &$ \Ket{-5}$            &$45.2$  &$ \Ket{-5}$            &$43.9$  &$ \Ket{-5}$            &$37.4$  &$\Ket{-5}$             &$36.3$  &$\Ket{-5}$             &$36.3$  &$\Ket{-5}$             &$36.3$  \\
					5     &$\Ket{0}$              &$119.3$ &$\Ket{0}$              &$114.9$ &$\Ket{0}$              &$102.4$ &$\Ket{0}$              &$101.5$ &$\Ket{0}$              &$101.5$ &$\Ket{0}$              &$101.5$ \\
					6     &$0.7(\Ket{4}-\Ket{-4})$ &$124.9$ &$0.7(\Ket{4}-\Ket{-4})$&$122.5$ &$\Ket{1}$              &$115.6$ &$\Ket{1}$              &$115.0$ &$\Ket{1}$              &$115.0$ &$\Ket{1}$              &$115.0$ \\
					7     &$0.7(\Ket{4}+\Ket{-4})$&$124.8$ &$0.7(\Ket{4}+\Ket{-4})$&$122.5$ &$\Ket{-1}$             &$115.6$ &$\Ket{-1}$             &$115.0$ &$\Ket{-1}$             &$115.0$ &$\Ket{-1}$             &$115.0$ \\
					8     &$\Ket{1}$              &$131.2$ &$\Ket{1}$              &$127.0$ &$0.7(\Ket{4}-\Ket{-4})$&$116.4$ &$0.7(\Ket{4}-\Ket{-4})$&$116.2$ &$0.7(\Ket{4}-\Ket{-4})$&$116.2$ &$0.7(\Ket{4}-\Ket{-4})$&$116.2$ \\   
					9     &$-\Ket{-1}$            &$131.2$ &$-\Ket{-1}$            &$127.0$ &$0.7(\Ket{4}+\Ket{-4})$&$116.4$ &$0.7(\Ket{4}+\Ket{-4})$&$116.2$ &$0.7(\Ket{4}+\Ket{-4})$&$116.2$ &$0.7(\Ket{4}+\Ket{-4})$&$116.2$ \\  
					10    &$0.7(\Ket{2}-\Ket{-2})$&$155.1$ &$0.7(\Ket{2}-\Ket{-2})$&$151.2$ &$0.7(\Ket{2}-\Ket{-2})$&$142.6$ &$0.7(\Ket{2}-\Ket{-2})$&$142.6$ &$0.7(\Ket{2}-\Ket{-2})$&$142.6$ &$0.7(\Ket{2}-\Ket{-2})$&$142.6$ \\
					11    &$0.7(\Ket{2}+\Ket{-2})$&$155.5$ &$0.7(\Ket{2}+\Ket{-2})$&$151.6$ &$0.7(\Ket{2}+\Ket{-2})$&$143.0$ &$0.7(\Ket{2}+\Ket{-2})$&$143.0$ &$0.7(\Ket{2}+\Ket{-2})$&$143.0$ &$0.7(\Ket{2}+\Ket{-2})$&$143.0$ \\
					12    &$\Ket{3}$              &$164.5$ &$\Ket{3}$              &$159.3$ &$\Ket{3}$              &$153.0$ &$\Ket{3}$              &$153.3$ &$\Ket{3}$              &$153.3$ &$\Ket{3}$              &$153.3$ \\
					13    &$\Ket{-3}$             &$164.5$ &$\Ket{-3}$             &$159.3$ &$\Ket{-3}$             &$153.0$ &$\Ket{-3}$             &$153.3$ &$\Ket{-3}$             &$153.3$ &$\Ket{-3}$             &$153.3$ \\ \hline
				\end{tabular} \caption{Eigenfunctions $\Psi_{i}$ and eigenvalues $E_i$ as obtained from the model described above.}
			\label{eigenvectors}
			\end{minipage}	}	}
			\end{table}								
\newpage

\subsection{Neutron diffraction in Tb$_{0.6}$Y$_{0.4}$RhIn$_5$ } 

\begin{figure}[h!]
	%\begin{center}
	\includegraphics[%
	width=0.8	\linewidth,keepaspectratio]{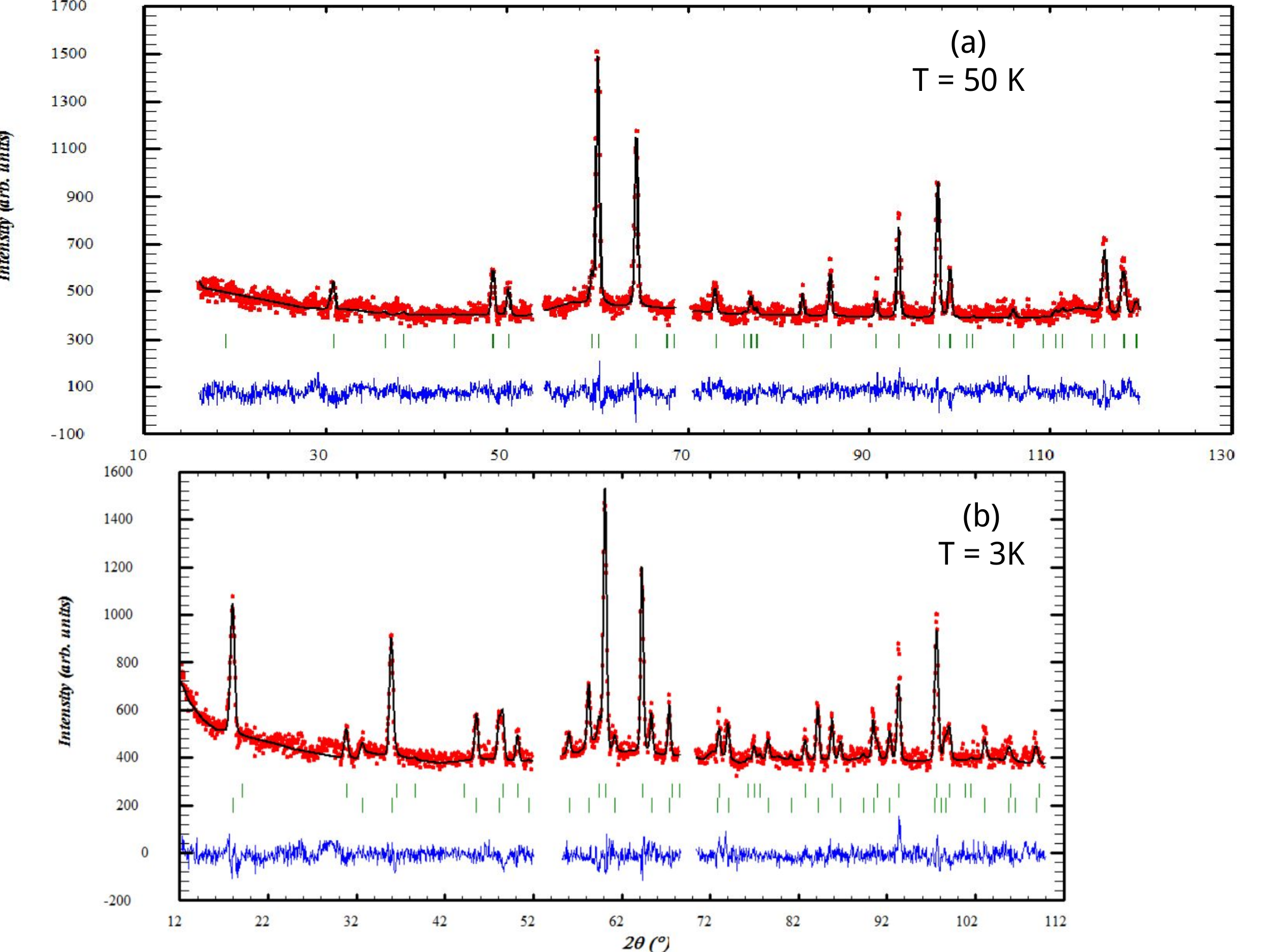}
	%\end{center}
	\caption{ Neutron powder diffraction patterns of Tb$_{0.6}$Y$_{0.4}$RhIn$_5$ obtained at 50 K (a) and 3 K (b).}
	\label{FigsNPD}
\end{figure}

\begin{figure}[h!]
	%\begin{center}
	\includegraphics[width=0.5\linewidth,keepaspectratio]{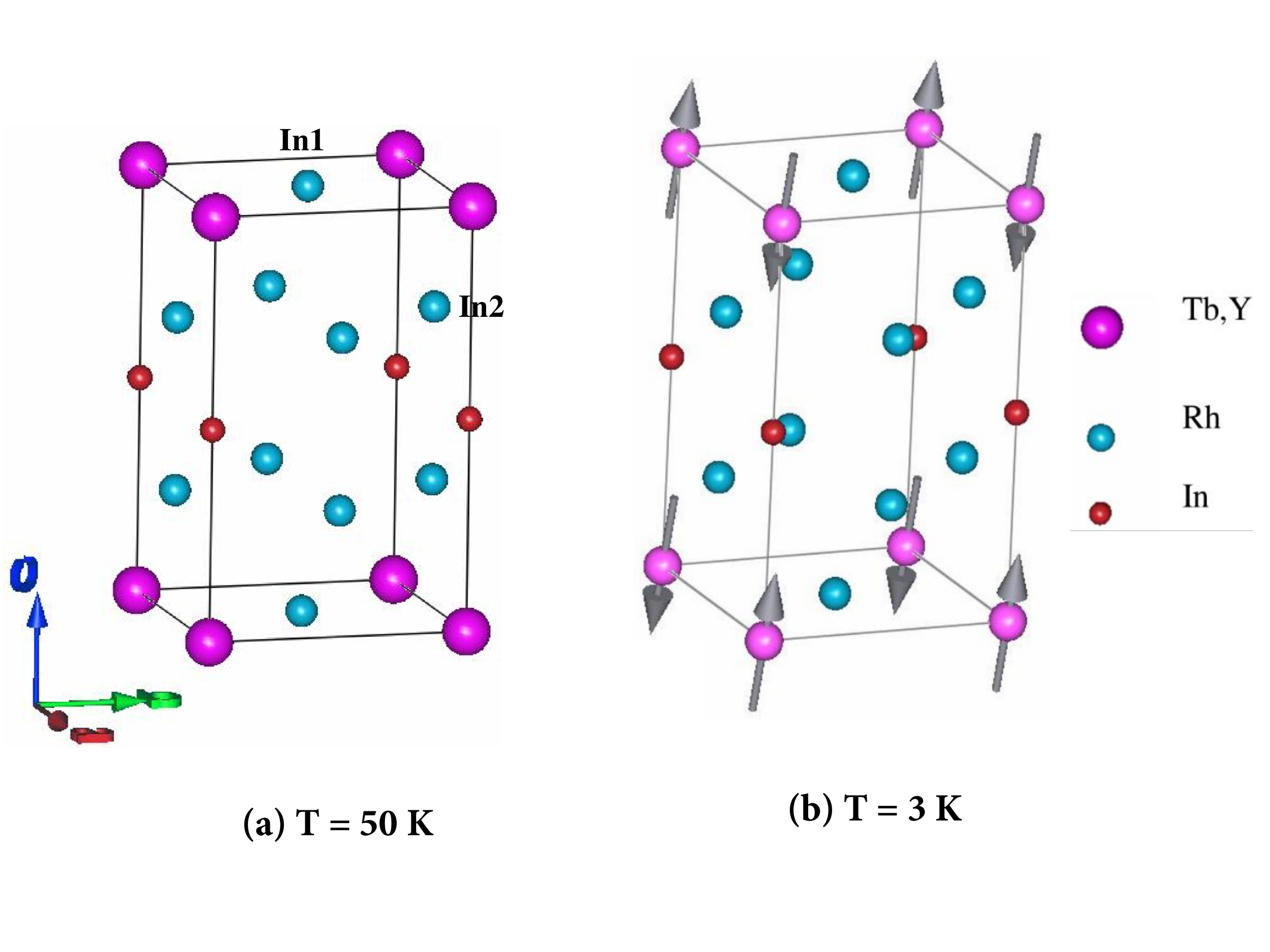}
	%\end{center}
	\caption{(Left panel) Chemical unit cell of the tetragonal Tb$_{0.6}$Y$_{0.4}$RhIn$_5$ (space group $P4/mmm$) above $T_N$; the Indium In1 and In2 sites are highlighted; (right panel) Magnetic unit cell according to RA for the Tb site. Arrows represent the total magnetic moment of Tb ions below $T_N$.}
	\label{UnitCell}
\end{figure}

We also investigated the microscopic magnetic structure for intermediate Y concentration by collecting PND data in the nominal Y-content sample Tb$_{0.6}$Y$_{0.4}$RhIn$_5$. Patterns at temperatures of 50 K and 3 K were obtained. The 50 K neutron pattern presented in Fig.~\ref{FigsNPD}(a) represents the nuclear scattering from the tetragonal $P4/mmm$ cell (Fig.\ref{UnitCell}). The refined lattice parameters at 50 K are $a=b=4.5835(5)$\AA , $c = 7.386(1)$\AA. The In atom occupy $4i$ sites in this $P4/mmm$ structure, generated by the special atomic position $(0~z~\frac{1}{2})$. The refined value of the positional '$z$' parameter is 0.3072(9). The reliability factors for this refinement are R-Factors: 4.32 and 5.36, $\chi^2$: 1.28, DW-Stat.: 1.2136, GoF-index: 1.1, Bragg R-factor: 16.2, RF-factor: 13.6. The refinement was made using a single nuclear phase. 

On the other hand, the diffraction pattern obtained at 3 K shows considerable magnetic contribution from the Tb AFM order [Fig.\ref{FigsNPD}(b)]. The propagation vector was found to be $k=[\frac{1}{2}~0~\frac{1}{2}]$. Also, Representational Analysis (RA) for the Tb site using the BASIREPS program (from the FullProf/WinPlotr suite) confirmed that the \textit{k}-vector $[\frac{1}{2}~0~\frac{1}{2}]$ is equivalent to $-k$. For this sample, at 3 K the Tb moment is $9.5 \pm 0.2 \mu_B$, with a canting angle of $(10 \pm 7)\degree$ off the \textit{c}-axis. The cell parameters are $a=b=4.5837(7)$\AA, $c= 7.379(2)$\AA~and the refined value of the \textit{z} position is $0.305(2)$\AA. The Bragg $R$-factor = 11.8, the $R_f$-factor = 9.03 and the magnetic $R$-factor = 12.3. 

The above results for neutron diffraction data was obtained by assuming that there is actually 0.68 Tb moles per formula unit. This agrees with the results from EDS spectrum (Fig.~\ref{FigEDS}) for this sample and with the fitting to hight T linear part of the inverse magnetic susceptibility, which points to an actual average composition between 0.67 and 0.7 moles of Tb.

It is worth noticing that we also collected PND data in the Tb$_{0.6}$La$_{0.4}$RhIn$_5$ compound (not shown), aimed to compare with the PND results of Tb$_{0.6}$Y$_{0.4}$RhIn$_5$. The analysis confirmed the magnetic structure previously studied by X-ray magnetic diffraction (XRMD) data\cite{raimundo5}, i.e. propagation vector $k=[\frac{1}{2}~0~\frac{1}{2}]$. However, the best fit to the data was obtained when we assumed a canting of Tb moments of $\sim$16$^{\text{o}}\pm 4^{\text{o}}$ from \textit{c}-axis, with a magnitude of 8.8 $\mu_B$ per Tb ion. For Tb$_{0.6}$Y$_{0.4}$RhIn$_5$ the best fit was obtained for Tb moments along \textit{c} and magnitude of 9.0 $\mu_B$ per Tb ion, which matches the expected value for Tb$^{3+}$ in the ordered phase. 

\section{Discussion}

\begin{figure}[h!]
	%\begin{center}
	\includegraphics[%
	width=0.7	\linewidth,
	keepaspectratio]{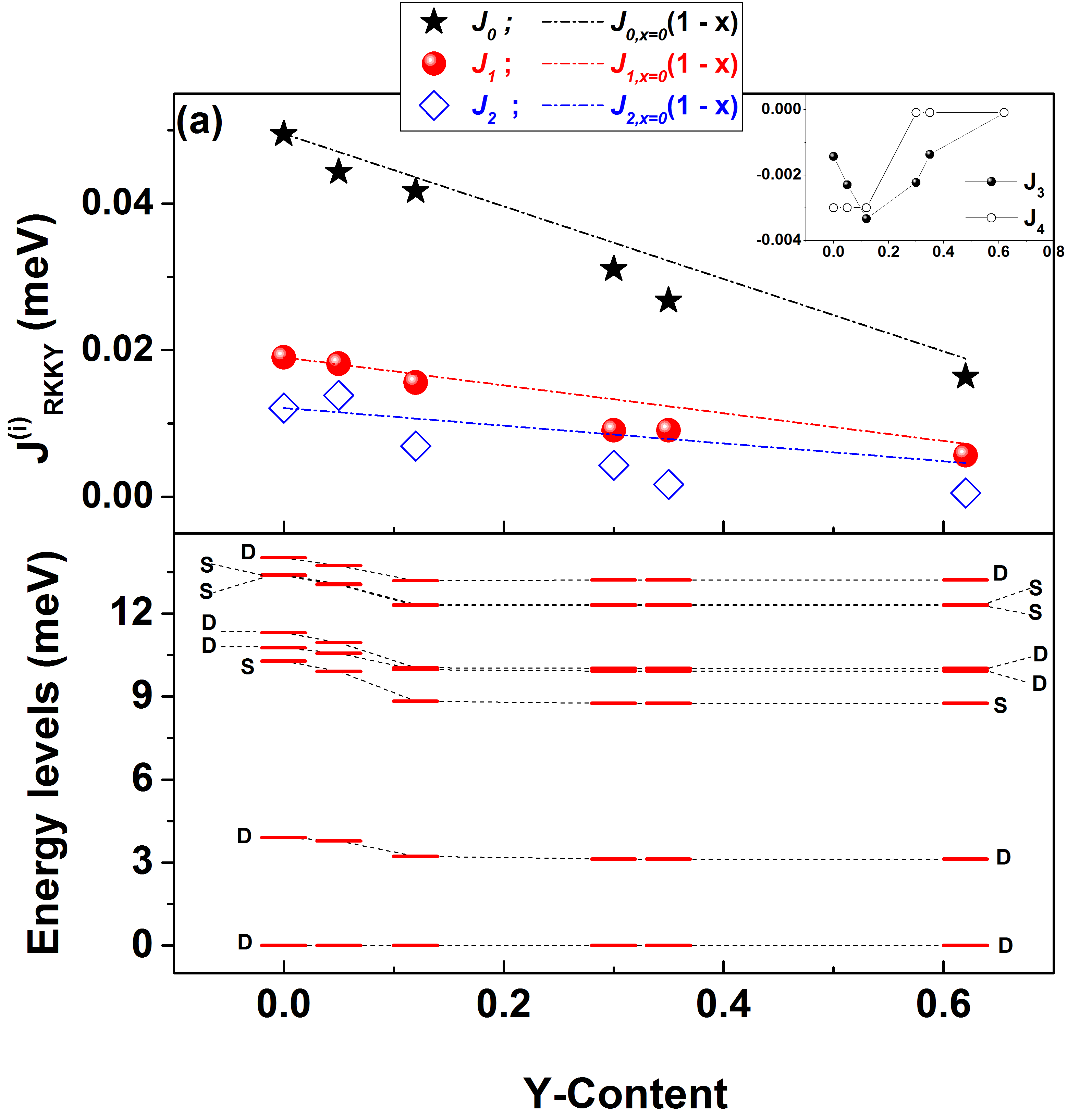}
	%\end{center}
	\caption{Top panel: Exchange parameters $j^{(0)}_{RKKY}$, $j^{(1)}_{RKKY}$, $j^{(2)}_{RKKY}$ vs. Y content. Dash-dotted lines show as would be a mean field behaviour (see text) for $j^{(0)}_{RKKY}$, $j^{(1)}_{RKKY}$, $j^{(2)}_{RKKY}$ the inset is the evolution of $j^{(3)}_{RKKY}$ and $j^{(4)}_{RKKY}$ vs. \textit{x}; bottom panel: Energy level schemes vs. \textit{x}. All parameters are in meV. 'S' stands for Singlet, and 'D' for Doublet. Dashed lines between different schemes are guide to the eyes.}
	\label{FigParamCEF}
\end{figure}

Our mean field simulations revealed that we achieved the best fits to the macroscopic data by using roughly the same crystal field parameters $B^{m}_{n}$ for all concentrations and varying only the exchange $j_{RKKY}^{(ik)}$ parameters. From Table \ref{resultados}, all $B_{0}^{4}$, $B_{4}^{4}$ and $B_{4}^{6}$ were kept fixed, while $B_{0}^{2}$ and $B_{0}^{6}$ have a maximum variation of $\sim$10\% with \textit{x}. Fig.~\ref{FigParamCEF}, top panel, shows the evolution of all the $j^{(0)}_{RKKY}$, $j^{(1)}_{RKKY}$, $j^{(2)}_{RKKY}$, $j^{(3)}_{RKKY}$ and $j^{(4)}_{RKKY}$ (inset). For $j^{(0)}_{RKKY}$, $j^{(1)}_{RKKY}$, $j^{(2)}_{RKKY}$ we also compared to a mean field behaviour line (e.g. $j^{(0)}_{RKKY}(x_i)= j^{(0)}_{RKKY}(x=0)\times(1-x)$). Both, $j^{(0)}_{RKKY}$ and $j^{(1)}_{RKKY}$ approximately follow the mean field decrease, which is not the case of $j^{(2)}_{RKKY}$, $j^{(3)}_{RKKY}$ and $j^{(4)}_{RKKY}$. A simple analysis of Tb-Tb ion distances along the directions where exchange parameters are defined, allow to see that for $j^{(2)}_{RKKY}$, $j^{(3)}_{RKKY}$ and $j^{(4)}_{RKKY}$, Tb ions are farther than for $j^{(0)}_{RKKY}$ and $j^{(1)}_{RKKY}$. A mean field decrease of the exchange is in agreement with the expected weakening of magnetic interactions with \textit{x} between first-nearest neighbors (FNN) in a  strongly localized system like the present one. This is roughly also the case for $j^{(1)}_{RKKY}$, which is not a FNN direction for Tb ions. Fig.~\ref{FigParamCEF}, bottom figure, shows the energy level distributions with the corresponding distribution of singlets (S) and doublets (D). The energy levels splitting due to the crystalline potential around Tb/Y ions is not expected to change with \textit{x} because of the close ionic radius of Tb and Y in the cuboctahedral environment. In fact, for a coordination number (CN) 12\cite{Kalychak} for \textit{R} ion in 1-1-5 compounds, there is atomic radii of 123 pm and 122 pm, respectively for Tb and Y,\cite{ShannonAtomicRadii,CN-RareEarth} while it is 136 pm\cite{ShannonAtomicRadii} for Lanthanum substituting Tb in $\mathrm{Tb}_{1-x}\mathrm{La}_{x}\mathrm{RhIn}_{5}$. For the latter, a CEF evolution with \textit{x} has been previously presented and discussed in ref.~ \onlinecite{raimundo5}. 

It is important to remark that when we compare the present fitting results with those obtained for $\mathrm{Tb}_{1-x}\mathrm{La}_{x}\mathrm{RhIn}_{5}$ compounds\cite{raimundo5}, the overall energy levels splitting proposed here is almost half ($\sim 14$ meV) the one calculated in refs. \onlinecite{raimundo2,raimundo5} for TbRhIn$_5$ ($\sim$26 meV and 30 meV, respectively). Further, N. V. Hieu et al.\cite{Takeuchi2007} got an overall splitting of about 19 meV, which is closer to the values obtained in this work. Even thought the definitive confirmation must come from inelastic neutron scattering and/or complimentary soft X-ray absorption\cite{HansmannXAS1,WillersXAS2} data, we should highlight an important point about the model used in the present work. Different from our previous reports, where we used an effective Tb-Tb isotropic exchange interaction term together with the $B^{m}_{n}$ parameters in the Hamiltonian\cite{pagliuso5}, this work uses a cluster of Tb ions with anisotropic first-neighbours RKKY interaction and the tetragonal CEF Hamiltonian (see section \ref{CEF}). This means that the complexity of magnetic exchange details are better captured with the present model. For instance, we cannot follow the details of the lower Y content (nominal \textit{x} = 0.15 and 0.3) experimental data with parameters obtained from the $\mathrm{Tb}_{1-x}\mathrm{La}_{x}\mathrm{RhIn}_{5}$ compounds data. On the contrary, we do followed the details of the latter with the parameters proposed for the former one. Worthwhile noted is that we also have conducted measurements of Linear Dichroism X-ray Absorption Spectroscopy (LDXAS) in single crystals of TbRhIn$_5$ in the $M_4$ and $M_5$ absorption edges of Tb ion, at the Brazilian Synchrotron Facility (LNLS), aimed to experimentally study the CEF ground state. However, the data collected is below the resolution needed to follow any change with linear polarization.

\begin{figure}[h!]
	%\begin{center}
	\includegraphics[%
	width=0.7	\linewidth,
	keepaspectratio]{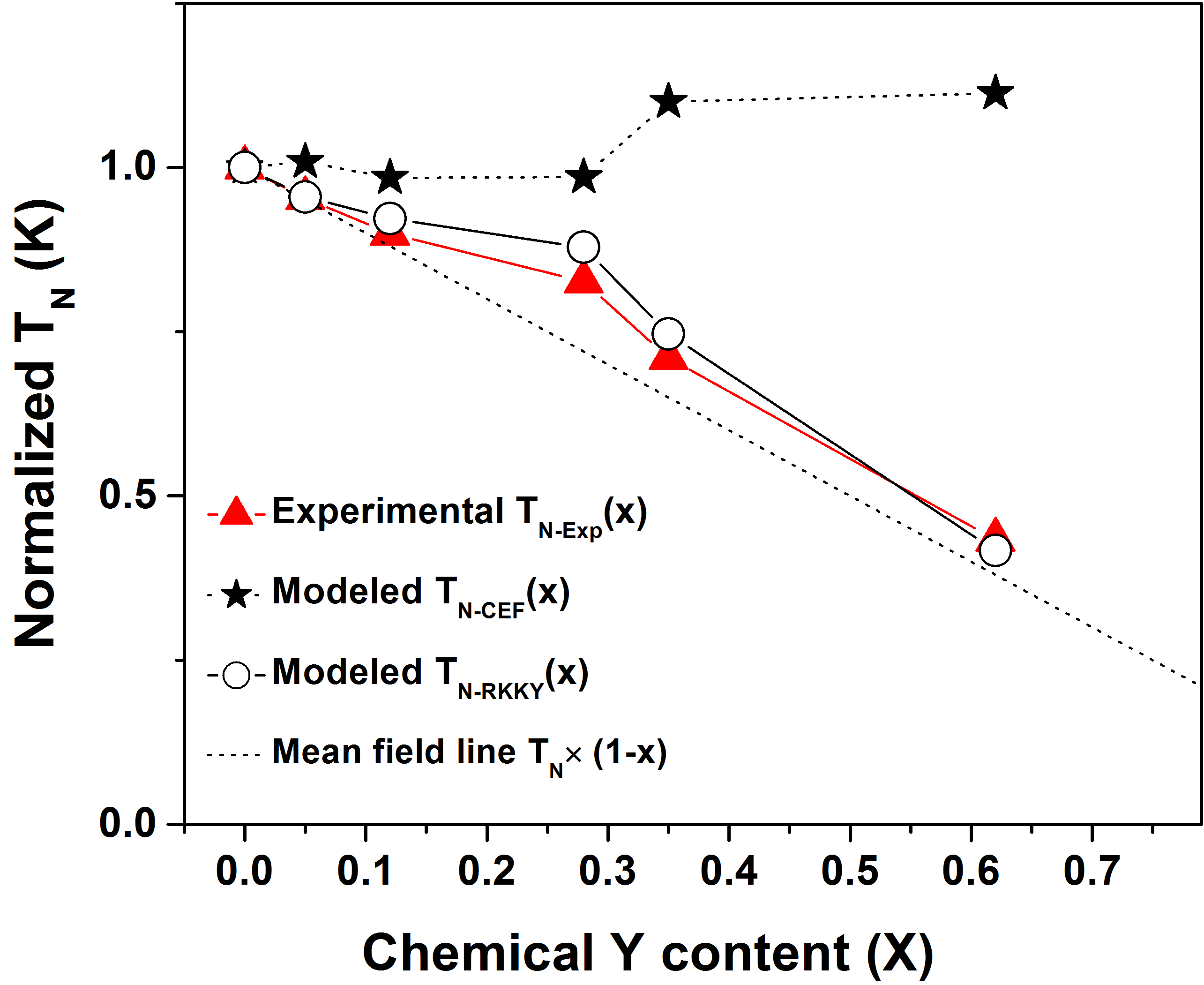}
	%\end{center}
	\caption{N\'eel temperature variation evolution with Y-content. Filled triangles represent the experimental variation of $T_N$ ($T_{N-Exp}$). Open circles are the variation of $T_N$ due to magnetic exchange weakening ($T_{N-{RKKY}}$). Closed stars signal the variation of $T_N$ due to the static electric potential from the environment ($T_{N-{CEF}}$). See text.}
	\label{TnBehavior}
\end{figure}

From the results of our simulations in the $\mathrm{Tb}_{1-x}\mathrm{Y}_{x}\mathrm{RhIn}_{5}$ family, we may infer that the evolution of N\'eel temperature and the magnetic structure with \textit{x} would be mostly the result of magnetic dilution effects. In Fig.~\ref{TnBehavior} we compare the experimental $T_N$ variation with Y-content by considering just that $j_{RKKY}^{(ik)}$ exchanges are changing, keeping the $B^{m}_{n}$s constant, and vice-versa, in our model. The experimental variation of $T_N$ ($T_{N-Exp}(x)$) was extracted from heat capacity data. For the variation of $T_N$ due to crystal field ($T_{N-CEF}(x)$), we simulated the data of compounds with $x\neq$ 0 ($x_i$), keeping constant the best results of $j_{RKKY}^{(ik)}$ from $x=$ 0. On the other hand, the change in $T_N$ due to changes in $j_{RKKY}^{(ik)}$s ($T_{N-RKKY}(x)$) is obtained by keeping constant the best $B^{m}_{n}$s from $x=$ 0 together with the best $j_{RKKY}^{(ik)}$ obtained for $x_i$ data. This comparison indicates that the experimental $T_N$ variation is accounted only by considering magnetic dilution effects, as expected. This is a simplified picture of a problem where we should consider the possible presence of substitutional disorder,\cite{raimundo8} frustration mechanisms or geometric fluctuations, mainly close to the critical concentrations.\cite{vojta3,StaufferPercolationTheory,Hoyos} For the present context, if the latter effects are present, they might be responsible for the separation of $T_{N-Exp}(x)$ from the mean field behaviour for intermediate concentrations.

Finally, we may comment on the canting of Tb magnetic moments from \textit{c}-axis observed in PND for Tb$_{0.6}$La$_{0.4}$RhIn$_5$ compound, as compared to the magnetic structures of TbRhIn$_5$ and nominal Tb$_{0.6}$Y$_{0.4}$RhIn$_5$. We argue that it could be related to changes in the strength of the crystalline field when La is changed by Tb. In particular, changes in the magnitude of $O_{2}^{0}=[3J_z^2-J(J+1)]$ and $O_{4}^{4}=\frac{1}{2}(J_{+}^{4}+J_{-}^{4})$, which are proportionals to the \textit{z} and \textit{ab}-plane projections, respectively, might be behind the observed canting. In fact, from our previous work on the Tb$_{1-x}$La$_{x}$RhIn$_5$ series we do observed the same propagation vector as for TbRhIn$_5$ $[\frac{1}{2}~0~\frac{1}{2}]$, but also important variations of the $B^{m}_{n}$ parameters with \textit{x}\cite{raimundo5}. Further, in ref.~\onlinecite{raimundo5} we could not determine the magnetic moment orientation with the available data. Therefore, the present PND data has allowed to conclude the magnetic dilution studies of these two series and enriched the understanding of crystal field effects along the $R$RhIn$_5$ family.

\section{Conclusion}

In this work we reported the results of the low temperature magnetic properties of Y-substituted TbRhIn$_5$ antiferromagnetic compound (for nominal Y concentrations $0.15, 0.3, 0.4, 0.5$ and $0.7$). Our magnetization and specific heat data were successfully simulated with a mean field model that accounts for anisotropic exchange and crystalline electric field excitations. From these results, we confirmed that the approximately same ionic radius of Y$^{3+}$ and Tb$^{3+}$ left unaltered the crystal field contribution  with \textit{x}, as somehow expected. For the latter, $T_N$ decrease is mostly determined by the weakening of the Tb$^{3+}$-Tb$^{3+}$ magnetic exchange. Neutron diffraction measurements in compounds with nominal compositions Tb$_{0.6}$La$_{0.4}$RhIn$_5$ and Tb$_{0.6}$Y$_{0.4}$RhIn$_5$ show that the AFM propagation vector is  $\vec{k}=[\frac{1}{2}~0~\frac{1}{2}]$, as determined for TbRhIn$_5$, with magnetic moments oriented along the tetragonal \textit{c}-axis for Tb$_{0.6}$Y$_{0.4}$RhIn$_5$ and canted magnetic moment ($\sim$16$^{\text{o}}$) from \textit{c}-axis in Tb$_{0.6}$La$_{0.4}$RhIn$_5$. The robustness of the TbRhIn$_5$ magnetic structure, and the relative strength between different $j_{RKKY}^{(ik)}$ parameters has been tested along both La and Y series. For La-substitution, both crystal field and magnetic dilution effects are relevant in the evolution of $T_N$ and magnetic structure. This work concludes a series of parallel magnetic dilution studies in Lanthanum substituted \textit{R}RhIn$_5$ compounds: GdRhIn$_5$,\cite{raimundo8} TbRhIn$_5$\cite{raimundo5} and NdRhIn$_5$\cite{raimundo7}. In the context of the series $R_{m}M_{n}$In$_{3m+2n}$ ($R=$ Ce-Tb; $M$ = Co, Rh or Ir; $m=1$ e $n=2$), this study confirms the relevance of CEF effects induced by changes in the crystal structures of diluted non Kondo tetragonal systems.

This work was supported by the FAPEMIG-MG (APQ-02256-12), CNPq (309647/2012­-6, 308355/2009-1, 2010-EXA020 and 304649/2013-9), FAPESP-SP (06/50511-8 and 12/04870-7). RLS is particularly grateful to CAPES Foundation (Brazil) for grant EST-SENIOR-88881.119768/\\2016-01.

\end{document}